\documentclass[runningheads]{svmult}
 
\usepackage{physprbb}  % modified textarea for proceedings,
\begin{document}
\title*{Can We Date Starbursts?}
\toctitle{Can We Date Starbursts?}
\titlerunning{Can We Date Starbursts?}
\author{Ariane Lan\c{c}on\inst{1}}
\authorrunning{A. Lan\c{c}on}
\institute{Observatoire de Strasbourg, 11 rue de l'Universit\'e,
  F--67000 Strasbourg, France}
 
\maketitle              % typesets the title of the contribution
 
\begin{abstract}
Age dating starbursts is an exercise with many caveats. We attempt to 
summarise a discussion session that was lead along a rather optimistic
guideline: the aim was to highlight that current age estimates, despite
undeniable uncertainties, do provide constraints on 
the physics of starbursts. In many cases, better starburst theories
will be needed before the improvement of empirical timelines
becomes crucial.
\end{abstract}

\section{Introduction} 
Many questions can be asked about our ability to trace the history of 
star formation (SF) in starbursts. The phrasing chosen by the Organizing
Committee of this workshop was \lq \lq Can we date starbursts\,?".
This formulation calls for one of only two answers\,: {\em yes}, or {\em no}...
When hearing the question, one automatically recalls
ones most recent conversation about the complexity of starburst
galaxies or about uncertainties in stellar population synthesis models.
Is there any chance for a positive answer\,? In preparing  guidelines
for the discussion, we took the optimistic approach of attempting to
defend a {\em yes}. Of course the final answer ended up not being 
as clear-cut, but some negative intuitions were countered.

Clearly, our degree of satisfaction with starburst age or duration 
measurements depends on the intended application. The initial 
question really holds two: how accurately and reliably can we date
starbursts\,? and is that sufficient to make astrophysics
progress\,?

Starburst galaxies are composite objects. The SF may occur
both in a diffuse mode and in clusters \cite{Metal95}.
The global duration of active episodes can approach $10^9$\,yrs,
while individual starburst clusters are often thought to
form instantaneously ($<10^6$\,yrs). To avoid confusion in the 
meaning of the word \lq \lq starburst", the following pages deal 
successively with (i) individual young starburst clusters, (ii)
individual intermediate age \lq \lq post-starburst" clusters that
trace starburst activity of the recent past, and (iii) starburst
galaxies as a whole. More extensive reviews and references regarding
the age dating of stellar populations can be found in 
\cite{Crete96}, \cite{Annapolis99}, \cite{Strasbourg00}
and in this volume.

\section{Individual young clusters}

This section focuses on starburst clusters with ages below $10^7$\,yrs,
as observed in large numbers in the main body of starburst galaxies
(\cite{GS99}, \cite{WZLetal99}, \cite{Boek00})
or in tidal tails of interacting objects (\cite{WZLetal99},
\cite{GHCZ00}). 

The conditions for cluster age determination are most favourable  
when the spectroscopic study of individual stars is possible. 
Until now such studies have been limited to the local neighbourhood
of the Milky Way, where many young OB associations exist but massive
compact young clusters (as seen in starburst galaxies) are rare/non-existent;
30\,Doradus in the LMC and NGC\,3603 in the Milky Way are
the most relevant accessible targets.
Nevertheless, the nearby objects highlight some of the difficulties:
\begin{itemize}
\item Samples of cluster stars with spectroscopically confirmed positions in
 the Hertzsprung-Russell (HR) diagram are small and strongly affected by
 stochastic fluctuations or spatial variation in the extinction; 
 they are potentially contaminated by field stars.
\item Massive star main sequence lifetimes vary between authors by up to 
 $\sim$\,25\,\%.
\item Rotation is poorly understood, but rotational velocities above 
 100\,km/s are the rule in early type stars. Meynet (in \cite{Annapolis99})
 shows that the main sequence lifetime of a massive star may be extended
 by 20--25\,\% in case of rotation. 
 % This directly affects age estimates.
\item The proportion of double stars and the effect of binarity on evolution
 are unknown. Binaries are usually neglected in predictions of 
 frequently used properties such as the number fractions of various types of 
 Wolf-Rayet stars.
\end{itemize}
In more distant starburst clusters, one integrates the
cluster light. The photospheric and wind features in the UV spectrum
are considered the most sensitive age indicators and
in principle give instantaneous burst ages to within a few Myr 
\cite{LSGetal99}. The study of line equivalent widths allows similar 
formal age accuracies if the light of the
whole H\,II region surrounding the cluster can be summed, the fraction
of escaping Lyman continuum photons considered negligible and the 
continuum contamination by background stars subtracted.
The above-mentioned problems associated with rotation, 
binarity and stellar tracks remain. 
Charlot (in \cite{Crete96}) for instance points out a delay of about
0.1\,dex (25\,\%) between the appearance of the first red star
contributions in two sets of commonly used evolutionary tracks.
The risk of stochastic fluctuations between
the properties of clusters with identical ages also persists
because of fluctuations in the 
small numbers of very luminous stars. Monte Carlo simulations 
\cite{CLC00} indicate that these fluctuations 
contribute less than $\sim$1\,Myr additional uncertainty to the age 
estimate as long as clusters more massive than $10^4$\,M$_{\odot}$
are considered. 

How the described sources of uncertainties add up or compensate 
each other is not known.  Today, if telescope time is not a limiting factor,
a detailed multi-wavelength study of a young cluster can be 
thought to provide an age 
estimate to better than $\sim$50\,\%. Opinions in the workshop audience 
varied from 30\,\% (which I would support at least in favourable cases), 
to a provocative 0.3\,dex (which are probably realistic at extreme metallicities
or in environments of particularly complex structure).

Can astrophysical questions be addressed with a 50\,\% accuracy in young
starburst cluster ages\,? Problems of physical interest include
SF processes themselves (delay between an external trigger and
the onset of SF, formation timescale for massive clusters,
propagation of SF within a galaxy) and their effects on
the environment (survival times of molecular clouds around
starburst clusters, bubble expansion timescales). 

Many examples illustrate that spectrophotometric ages, despite the 
uncertainties, provide interesting constraints. 
%The resistance of dense molecular clouds against destruction by newborn stars 
%has been shown to exceed early expectations (Solomon? this workshop).
Age spreads of several Myr have been found in OB associations
(\cite{Wetal96}, \cite{CC98}, \cite{PS00}), showing
that a unique number does not suffice to describe their age.
WC/WN star number ratios indicate that spreads of a few Myr
may also be relevant to clusters in starburst environments \cite{SCK99}.
Rather complex age structures are seen in NGC\,3606 and 30\,Dor. 
In both cases the massive stars of the youngest,
2-3\,Myr old component, are concentrated in the central few parsecs and
surrounded with significantly older components (\cite{GBC99},
\cite{SMBT99}). This situation remains to be convincingly explained by 
cluster formation models (what are the relative roles of a progressive
onset of SF \cite{PS00}, mass segregation \cite{Meyl00}, \cite{Krou00}, 
propagation, merging\,?). Oey \& Massey \cite{OM95} studied the
LH 47/48 and the surrounding superbubble in the LMC, and found a significant
disagreement between the stellar ages and the bubble properties
predicted from a simple dynamical model, calling for more detailed
modelling of the reactions of the ISM. Uncertainties in the identification of 
external triggers and in {\em their} onset time dominate in many studies of the 
initiation of SF in young clusters. Clearly, spectrophotometric dating  has
been successful in providing other fields of starburst cluster research with
new problems.

\section{Individual post-starburst clusters}
Star clusters with ages between $10^7$ and $10^9$\,yrs are useful to relate
current SF activity to potential starbursts of the recent past. They are
often found together with the young clusters discussed previously. As
they do not ionize their surroundings and have already faded at
optical wavelengths, they have not yet been searched for and studied
as systematically as their younger counterparts.

Post-starburst clusters are dominated by B then A type stars in the optical/
near-UV, by red supergiants (RSG) and then giants of the upper asymptotic
giant branch (TP-AGB) in the near-IR. The effects of mixing processes,
due e.g. to rotation, appear essential to explain the location of B stars in
the HR diagram \cite{Len99}; Figueras \& Blasi \cite{FB98} use simulations
of the Str\"omgren photometry of stellar populations with reasonable
rotation velocity distribution to conclude that photometric ages 
are affected at the 30-50\,\% level. More consistent approaches
combining the effects of rotation on internal structure and 
on observable properties have not yet been systematically
applied to age studies. Supergiant counts should be used with caution
at non-solar metallicities (Z) as the Z-dependance of the blue/red
number ratio is not predicted correctly by models \cite{LM95}. It seems 
that at Z$_{\rm M31} \simeq \rm{Z}_{\odot}$ the RSGs have later spectral 
types but are only produced for $m<15\,{\rm M}_{\odot}$ (age $\sim$12\,Myr) as
opposed to $m<25-30\,{\rm M}_{\odot}$ (age $\sim$7\,Myr) at 
Z$_{\rm NGC6822}\simeq \rm{Z}_{\odot}/3$ \cite{Massey98}. 
Modelling the thermal pulses and
the Mira-type pulsation along the TP-AGB, in addition to the 
early AGB, is essential when studying stars in $10^8-10^9$\,yr old clusters.
Number counts that separate C-rich stars from O-rich stars of various 
subtypes then are potential age-indicators (Lan\c{c}on \& Mouhcine, this
volume, and references therein).

For unresolved solar metallicity clusters younger than $\sim 50$\,Myr, 
well-isolated from the host galaxy background, the UV features give ages to 
within $\sim 20$\,\% \cite{dMLH00}. Effects of metallicity are uncertain,
but empirical calibrations are being attempted (Tremonti, this workshop). 
The absorption line spectrum (H{\sc i} and metals)
{\em together} with the energy distribution in the Balmer region gives 
ages to within $\pm 30\,\%$ (\cite{GS99},
\cite{GDLH99}). Reddening-independent colour-indices
%\`a la Becker (1938), as used for the NGC\,4038/39 clusters 
in \cite{WZLetal99} are efficient and could be 
generalised to include  near-IR fluxes. Gilbert (this workshop) showed that, 
at a given metallicity, near-IR spectra of synthetic clusters with ages of
$10-25$\,Myr (RSG-dominated) and age differences of $2-3$\,Myr can be 
distinguished and sorted. TP-AGB stars leave potentially useful 
spectral signatures in integrated spectra of slightly older objects
(\cite{LMFS99}). Stochastic fluctuations in 
the integrated spectrophotometry, that are dominated by the most luminous red 
stars, add negligible amounts to the other dating errors as long as the 
clusters contain more than $10^4$\,M$_{\odot}$ of stars \cite{LM00}.

Again, when enough telescope time can be obtained to combine several of the 
above approaches, ages can be expected to within a conservative 
$\pm 50\,\%$ (25\,\% in favourable cases, 0.3\,dex for sceptical attendees).
 
The ages discussed here are comparable to galaxy interaction 
timescales and more generally to the duration of 
starburst activity on galaxy scales. Mihos (this workshop)  
reminded us that the treatment of the transition from a dynamical 
perturbation to star formation in dynamical models is simplistic; delays
of 100 to 500\,Myr are found to be typical before onset of starburst activity.
Obvious morphological signatures of an interaction
fade away over similar timescales; in the case of NGC\,4038/39
the spectrophotometric age distribution of the clusters is probably a
safer indication of a second encounter than model adjustments to the
projected system structure. In NGC\,1614 and IC\,342 (Rieke, Genzel, this 
session) starburst knots form a $\sim$0.5\,kpc nuclear ring, with younger 
knots (H$\alpha$ sources, $\leq 6$\,Myr old) located at larger galactic 
radii than older ones (RSG hosts, $\leq 7$\,Myr old). No 
dynamical models are as yet available to explain this situation well
enough to require improved spectrophotometric ages. 

Is the formation of generations of starburst clusters a recurrent
phenomenon? When cluster ages become comparable to the dynamical
timescales of a galaxy, age differences much shorter than this 
time cannot be interpreted as separate SF episodes, but rather as one extended
one. Therefore a 50\,\% precision on the age is sufficient to detect
potential separate episodes. Then, attempts to the compare properties
of the starburst clusters of the current and the previous active phases
must deal with a large variety of dynamical effects that rule
the survival/destruction of starburst clusters over timescales
of $10^8-10^9$\,yrs \cite{Ger00}. Uncertainties in those are likely
to wash out 50\,\% age errors.

In this section again, our (biased) approach demonstrates that
current age estimates pose challenging astrophysical
problems that are far from being resolved to the point of necessitating
better timelines.

\section{Starburst galaxies at low spatial resolution}

Let us finally question the dating of starburst galaxies
observed at a spatial resolution no better than a few 100\,pc, or completely
unresolved. Partial spatial resolution has obvious advantages but also 
has some dangers: aperture mismatch between wavelengths, the
likelihood that wavelength-dependent photon-exchanges
with regions outside the line of sight (through scattering) falsify
energy balances, the possibility that average obscuration curves 
don't apply, etc. The youngest and/or least reddened
stellar component is usually dominant at UV wavelengths; but
underlying \lq \lq evolved" populations have been found in all
starbursts. Age studies must also aim at determining whether these
are part of an extended starburst episode that is still going on, or whether
they are remnants of previous, dynamically unrelated star formation.

The nuclear starburst in the interacting spiral galaxy NGC\,7714  will
be used here for illustration. Integrated photometry is available over the
whole electromagnetic spectrum. Extinction is very inhomogeneous and
typically $A_v \sim 0.8$. A recent study \cite{LGLG00} addresses
the photometry and the UV+V+near-IR spectra of the central 300\,pc.
There, the UV is dominated by a young ($\sim$\,5\,Myr old) burst, obviously
seen through a hole in the dust distribution; the short wavelengths thus
contain no information on putative other young populations, including
those required to explain the far-IR emission. The broad band photometry
can be adjusted satisfactorily with many 
models: continuous SF over as little as a few $10^7$\,yrs or as long 
as $\sim 10^9$\,yrs, or a succession of brief bursts: dust distributions
provide more than enough degrees of freedom. 
%The global constraints provided by radio, X\,ray and far-IR fluxes
%(not available through directly comparable apertures) can be accommodated
%to within the uncertainties. 
More stringent constraints come from 
spectroscopy\,: the Hubble Space Telescope UV spectrum favours the 
presence of at least one instantaneous 5\,Myr burst\,; the Balmer line
region rejects the optical predominance of populations younger than 
$\sim 300$\,Myr or older than $\sim 900$\,Myr (note that the continuum shape 
had to be used in addition to the line profiles of the rectified spectrum
in order to reach this conclusion); the K band spectrum suggests
mixed contributions, as opposed to a population purely dominated by   
RSG or by TP-AGB stars. The far-IR flux sets a loose upper limit on the 
amount of heavily obscured young stars, and the reddened Balmer ratio 
a lower one. The study concludes that starburst activity has been going on 
with ups and downs over an extended time, and that durations between $\sim 300$
and $\sim 900$\,Myr are consistent with the data. This is an 
age to $\pm 50\,\%$.

The observational constraints on starburst studies can and must 
still be improved,
using available instruments; but on the other hand, more dust 
configurations and the effects of chemical evolution must be explored
systematically, adding even more free parameters. We will thus probably
have to bear with $\pm\,50\,\%$ estimates for a while.

Is that enough\,? In the case of NGC\,7714, it is at least sufficient
to point out an astrophysical problem\,: the comparison of the system 
morphology with dynamical simulations indicates that the
closest encounter with 
with NGC\,7715 occured about 100\,Myr ago. The model parameters would allow to 
increase the time since interaction by about a factor of 2, but it seems 
difficult to reconcile this dynamical timescale with the starburst 
timescales derived from spectrophotometry. 

\section{Conclusion}

Although some workshop participants never accepted 
age uncertainty estimates below 0.3\,dex, we believe that
detailed multiwavelength studies, as possible with current instruments
(when access to them is not a limiting factor), allow to reach
$\pm 50\%$, or even better in particularly favourable configurations.
The session has allowed many examples to be discussed, and we
hope it has conveyed the positive impression that current age
determinations, despite their uncertainties, are indeed providing 
essential constraints on theoretical issues related to starbursts.

%INDEX%%%%%%%%%%%%%%%%%%%%%%%%%%%%%%%%%%%%%%%%%%%%%%%%%%%%%%%%%%%%%%%
% Please check with the editor of your book whether he plans to
% include a "mutual" subject index - if so, please code your entries
% in the standard syntax. For your own purposes you may print your
% "personal" index by using the following commands:
%
%\clearpage
%\addcontentsline{toc}{section}{Index}
%\flushbottom
%\printindex
%%%%%%%%%%%%%%%%%%%%%%%%%%%%%%%%%%%%%%%%%%%%%%%%%%%%%%%%%%%%%%%%%%%%%

\end{document}